\documentclass[epj-spec]{svjour}
\usepackage{graphics}
\usepackage[]{amsmath}
\usepackage[]{amssymb}
\newcommand{\mb}[1]{\mbox{\boldmath $#1$}}
\newcommand{\mat}[4]
{
\left(
\begin{array}{cc}
#1 & #2 \\
#3 & #4 
\end{array}
\right)
}
\newcommand{\mvec}[2]
{
\left(
\begin{array}{c}
#1  \\
#2  
\end{array}
\right)
}
\newcommand{\mib}[1]{\mbox{\boldmath $#1$}}
\newcommand{\bfk}{\mb{k}} 
\begin{document}
\title{Topological aspects of graphene }
\subtitle{ Dirac fermions and the bulk-edge correspondence in magnetic fields
}

\author{
Y. Hatsugai
\inst{1} 
\and 
T. Fukui 
\inst{2} 
 \and
H. Aoki
\inst{3} 
}
\institute{
Department of Applied Physics, University of Tokyo, Hongo, Bunkyo-ku, Tokyo 113, Japan
\and
Department of  Mathematical Sciences, Ibaraki University, Mito 310-8512, Japan
 \and 
Department of Physics, University of Tokyo, Hongo, Bunkyo-ku, Tokyo 113, Japan
}
\abstract{
  We discuss topological aspects of electronic properties of graphene, 
including edge effects, with the tight-binding model
 on a honeycomb lattice and its extensions 
to show the following: 
(i) Appearance of the pair of massless Dirac dispersions, which is the 
origin of anomalous properties including a peculiar quantum Hall 
effect (QHE), is not accidental to honeycomb, but is rather 
generic for a class of two-dimensional lattices that interpolate 
between square and $\pi$-flux lattices.  Persistence of the 
peculiar QHE is interpreted as a topological stability.
 (ii) While we have the massless Dirac dispersion only around $E=0$, 
the anomalous QHE associated with the Dirac cone 
unexpectedly persists for a wide range of the chemical potential.  
The range is bounded by van Hove singularities, at which we predict 
a transition to the ordinary fermion behavior accompanied by 
huge jumps in the QHE with a sign change.  
(iii) For edges we establish a coincidence between the quantum Hall effect 
in the bulk and the quantum Hall effect for the edge states, 
which is a manifestation of the 
topological bulk-edge correspondence.  We have also explicitly 
shown that the $E=0$ edge states in honeycomb in zero 
magnetic field persist in magnetic field.  
} 
\maketitle
\section{Introduction}
\label{intro}
One fascinating aspect of the condensed-matter physics is that 
we can have various field theories effectively realized on 
low-energy scales.  Electronic structure in a honeycomb lattice 
is of particular interest, since the system 
has 
zero-mass Dirac cone dispersion at Brillouin zone corners, 
so we have got an intriguing mixture of a continuum 
limit and the lattice effect.  
Graphene, a single layer carbon atoms, provides
a prototypical realization of characteristic electronic structures of 
a honeycomb lattice.  This 
provides interesting problems
in condensed matter physics especially in its topological aspects, 
which is the subject-matter of the present article.
Its band dispersion is composed of a pair of $k$-linear 
bands that touch with each other around $E=0$,  
so that the graphene is a
condensed-matter realization of zero-mass Dirac fermions 
around $E=0$.
Now, the Dirac particles have been known to provide  topologically non-trivial  
structures\cite{Seme84,Hal87}.
Physics on the honeycomb lattice thus provides potential links 
between the field-theoretic
concepts and various  condensed matter phenomena
\cite{KM05,Lud93,Hat90}.

Indeed, recent experiments have revealed a peculiar type 
of quantum Hall effect\cite{Gus05,Nov05,Zha05}.
Graphite itself has been studied 
extensively from many different points of view.  
There has also been a substantial 
amount of development as  basic materials 
in nano-technologies\cite{ZhengAndo,Ando05}, where 
a special interest is  
characteristic 
boundary magnetic properties 
which originate from edge states\cite{Fuji96,Waka98}.
The fact that the honeycomb lattice as interesting 
edge properties as well as interesting bulk properties 
is no accident, but this is in fact a realization of
hidden physical structures as a topological order\cite{Wen89,Hat04,Hat05}.

Topologically non-trivial ground states  are, 
unlike ordinary symmetry-broken states, characterized 
by geometrical phases rather than by local order parameters.  
Geometrical phases which are fundamental in quantum physics 
provide basic tools for characterizing
topological orders\cite{Hat04,Hat05,Shapere89}.
Appearance of the edge states is one of the important physical 
 consequences of the non-trivial  
topological orders\cite{Laughlin81,Halperin82,Hatsugai93a,Hatsugai93b,yh-qhe-review}.
Physical properties of the bulk go hand in hand with those of the edge 
in the topologically ordered states.
So topological properties, which may not so apparent in the bulk, may  
emerge around the edges or defects with physical consequences.
\cite{Laughlin81,Halperin82,Hatsugai93a,Hatsugai93b}.
This bulk-edge correspondence 
is shared by various topologically ordered states,
where a typical one is the quantized Hall (QH) effects
\cite{Hatsugai93a,Hatsugai93b}.                      
Among various 
topologically ordered states, 
edge states of the Dirac particle is interesting, 
especially when the particle is 
massless\cite{Wit82,Niem83}.
This is in fact directly related with the edge states in graphene.
To be more precise, honeycomb has a bipartite lattice structure 
(i.e., a chiral symmetry), which 
 has a fundamental importance in the existence of the $E=0$ edge states.
From the topological point of view, 
appearance of  the magnetic moments 
in the edge states in graphene is understood topologically 
as a chiral symmetry breaking in edge states\cite{Ryu02,Ryu03cb}.

In this article,
 we discuss results obtained in the reference \cite{HFA06}
 focusing on the topological aspects of graphene.

\section{Topological Stability of Dirac Fermions}

\begin{figure}
\begin{center}
\resizebox{0.40\columnwidth}{!}{%
  \includegraphics{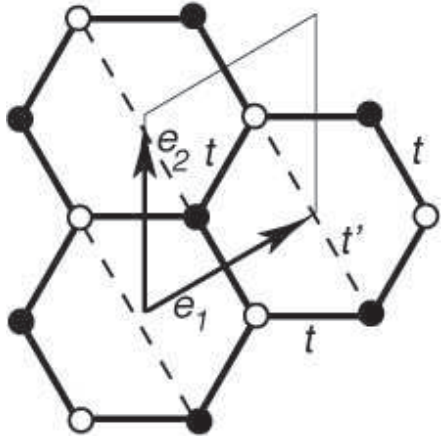} }
\caption{
A honeycomb lattice with a unit cell 
and the diagonal transfer $t'$ indicated. 
}
\label{fig:1}       
\end{center}
\end{figure}

We consider 
a tight-binding Hamiltonian, $H$,  for a honeycomb lattice
with nearest-neighbor hopping $H'$
\begin{alignat*}{1} 
H &=   H_{\rm honeycomb}+H'
\end{alignat*} 
\begin{alignat*}{1} 
H_{\rm honeycomb}  &= 
 t\sum_{\mb{j}} \Big[
c ^\dagger _\bullet(\mb{j})c_\circ(\mb{j})
+e^{i2\pi \phi j_1}c ^\dagger _\bullet(\mb{j})c_\circ(\mb{j}-\mb{e}_2)
+c ^\dagger _\bullet(\mb{j}+ \mb{e}_1)c_\circ(\mb{j})
\Big] +{\rm h.c.},
\\
H' &= 
t'\sum_{\mb{j}}e^{-i2\pi \phi (j_1+1/2)}
c^\dagger_\bullet(\mb{j}+\mb{e}_1-\mb{e}_2)c_\circ(\mb{j})+{\rm h.c.}
\end{alignat*} 
where $t=-1$ is the transfer energy (taken to be the unit of energy).  
Since the honeycomb, a non-Bravais lattice with two sites per unit cell 
is bipartite, we have defined two kinds of fermion operators, 
$c_\circ(\mb{j})$ and $c_\bullet (\mb{j})$ as in Fig. \ref{fig:1}, 
where $\mb{j}= j_1\mb{e}_1+j_2\mb{e}_2$ labels the unit cell 
with two translation vectors $\mb{e}_1 = (3/2 ,\sqrt{3}/2)a$ and
$\mb{e}_2 = (0,{\sqrt{3}} )a$.  

We then apply an external magnetic field, which is characterized by 
\[
\phi = BS_6/(2\pi) \equiv p/q,
\]
the magnetic flux 
penetrating each hexagon of 
area $S_6=(3\sqrt{3}/2)a^2$.  

Let us first discuss the energy dispersion in zero magnetic field.
 As is well known, 
the spectrum of the Hamiltonian $H_{\rm honeycomb}$ 
consists of two bands touching with each other 
at two points (K and K') in the Brillouin zone,
which are understood as doubled Dirac fermions.
Around the zero-gap points 
the energy dispersion is generically linear 
 in $(k_x, k_y)$ space.
 
Although the appearance 
of the massless Dirac fermions may seem just an accident 
for the honeycomb lattice, we shall show here 
that the appearance of massless 
Dirac fermions is, surprisingly enough, a generic property and 
stable against e.g. introduction of  
a diagonal hopping $H'$ if we preserve the chiral symmetry
(bipartite lattice with  
every transfer between A and B sites).
In momentum representation, the Hamiltonian in zero magnetic field
is given as
\begin{alignat*}{1} 
H &= \int \frac {d^2k}{(2\pi)^2} \mb{c} ^\dagger ( \mb{k} ) 
\mb{h} (\mb{k} ) \mb{c} (\mb{k})
\\
\mb{c} (\mb{k} ) &= \mvec{{c}_\bullet (\mb{k} )}{{c}_\circ(\mb{k} ) },
\\
\mb{h}(\mb{k} )=& 
t\mat{0}{\Delta(\mb{k} )}{\Delta^*(\mb{k} )}{0},
\\
\Delta( \mb{k} )
=& 1 + e^{-ik_2}+e^{-ik_1} \left[1 +(t'/t) e^{i k_2}\right] .
\end{alignat*}

\begin{figure}
\begin{center}
\resizebox{0.80\columnwidth}{!}{%
  \includegraphics{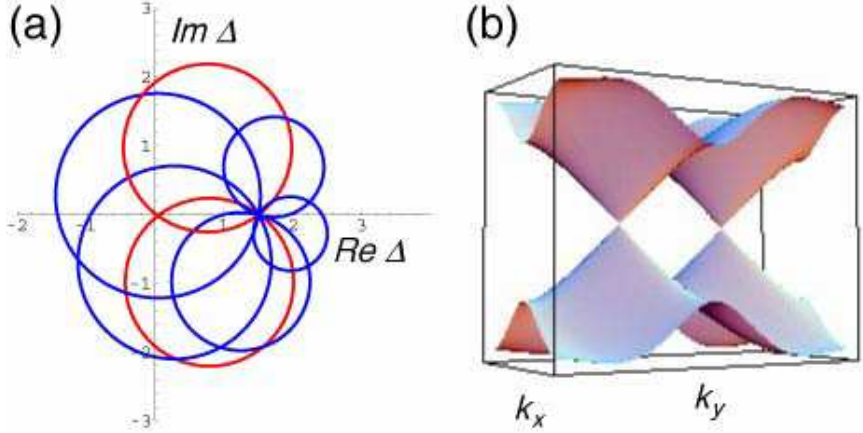} }
\caption{
(a) Locus (each circle) of $\Delta(\mb{k} )$
is shown on the complex $\Delta$-plane 
when we vary $k_x:0\to 2\pi$ for several fixed values of $k_y$ for $t'/t=-1/2$.
The red circle traverses the origin, which implies the appearance of
the gapless Dirac fermion. The doubling of the Dirac fermion 
is also clear from this geometric consideration.
(b) Corresponding energy dispersion for $t'/t=-1/2$.
}
\label{fig:2}       
\end{center}
\end{figure}
The chiral symmetry is expressed as
\begin{alignat*}{1} 
\{\mb{h} , \mb{ \gamma }\} &= 0,
\ \
\mb{ \gamma } = \mb{\sigma} _z=\mat{1}{0}{0}{-1},
\end{alignat*} 
which implies that the
energy bands are symmetric about $E=0$.  
For the honeycomb lattice 
$\Delta$ on the complex plane delineates a circle centered at 
$C_0=1 + e^{ik_2}$ with a radius 
$r=\sqrt{1+2(t'/t)\cos k_2+{(t'/t)}^2}$
when $k_1$ is varied from $0$ to $2\pi$ for a fixed value 
of $k_y$ (Fig.2).  
By examining $C_0$ and $r$, we can show that we have a 
Dirac cone when 
the circle goes through the origin, which occurs for $-3\le t'/t <1$.
This shows a topological stability for the appearance  
of the Dirac fermions. 
That is, Dirac cone persists when the lattice deviates from the 
honeycomb, even up to the limit of the square lattice where $t'/t=1$ 
in the present notation, unless the chiral symmetry is 
not broken. 
We can extend the present line of consideration to 
the general condition for the appearance of Dirac cones, 
which will be published elsewhere\cite{HFA07}.

\section{Hall conductance of the graphene as the Chern number}
The Hall conductivity of the noninteracting 
2D electron systems, as described with the 
linear response theory,\cite{Aoki81} 
may be regarded as a topological quantum 
number.\cite{Thouless82}  
Namely, when the Fermi energy lies in 
the $n$-th gap, the Hall conductivity is given by 
\begin{eqnarray*}
\sigma_{xy}=-\frac{e^2}{h} c_{\rm F}(E_{\rm F}),
\end{eqnarray*}
where $c_{{\rm F}}$ is an integer called the Chern number, 
which describes how the wave function
responds to a vector potential 
generated by Berry's gauge potential in the Brillouin zone.  
If we adopt a Bloch function in the $j$-th energy band
$\mb{\varphi}_j(\bfk)$ assuming that the energy gap opens 
over the entire Brillouin zone at each gap in the sequence    
$\epsilon_1(\bfk)\le\cdots
\le\epsilon_n(\bfk)<E_{\rm F}<\epsilon_{n+1}(\bfk)\le\cdots\le\epsilon_{2q}(\bfk)$.
We can then define a non-Abelian Berry's gauge potential, 
\begin{alignat*}{1} 
\mib{A} &= \mb{\Psi}^\dagger d  \mb{\Psi}  
\\
\mb{\Psi} &=  ( \mb{\varphi} _1,\cdots, \mb{\varphi} _n).
\end{alignat*} 
Here $A(\bfk)$ is an $n\times n$ matrix-valued one-form, 
which is a difference from the ordinary, scalar Berry's gauge potential.
Then the Chern number is given as 
the U(1) part of the U($n$) gauge potential,
\begin{eqnarray*}
c_{{\rm F}}(E_{\rm F})=\frac{1}{2\pi i}\int{\rm Tr}~d A.
\label{ConCheNum}
\end{eqnarray*}
This formulation, a non-Abelian 
extension of Berry's gauge potential,\cite{Hat04,Hat05} 
may seem too elaborate, 
but is useful when there are multiple bands below the Fermi energy.   
Namely, this formula holds {\it even if} 
some gaps in the Fermi sea 
are closed, as long as the bands in question, 
$\epsilon_n(\bfk)$ and $\epsilon_{n+1}(\bfk)$, 
do not overlap.  
In the special case when all the bands in the Fermi sea are 
separated with each other, 
the Chern number, Eq.(\ref{ConCheNum}), is simply the sum of 
the Chern numbers assigned to individual bands.

\begin{figure}
\begin{center}
\resizebox{0.55\columnwidth}{!}{%
  \includegraphics{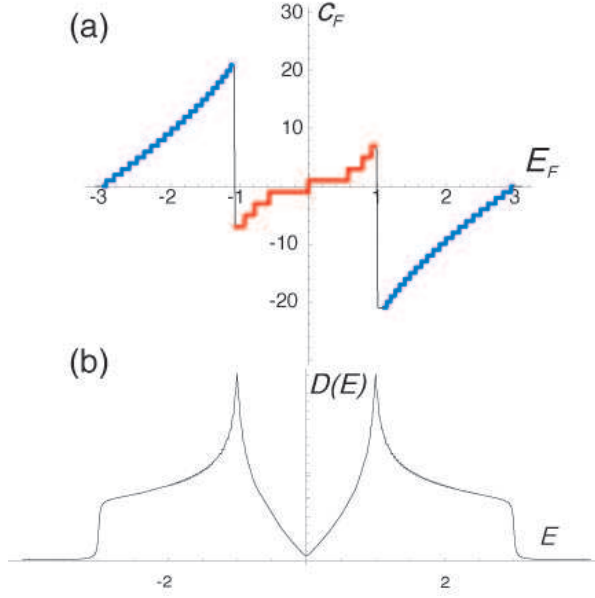} }
\caption{
(a) The Chern number (Hall conductivity per spin in units of $-e^2/h$) 
for magnetic field $\phi=1/31$ is plotted against 
the Fermi energy $E_{\rm F}$ for 
$t'/t=0$.
The red lines indicate the steps of two in the Chern number sequence 
($\sigma_{xy}=\pm(2N+1)e^2/h, N$: integer), 
while the blue lines step of one ($\sigma_{xy}=\pm Ne^2/h$).
(b) Density of states for the honeycomb lattice.}
\label{fig:3}       
\end{center}
\end{figure}

To evaluate the expression, here we have employed a technique developed recently 
in the lattice gauge theory by two of the present authors\cite{Fukui05,Hat05ep2ds}.
The numerical results 
for the Chern number $c_{\rm F}$, i.e., the 
Hall conductivity per spin,  
as a function of the chemical potential $E_{\rm F}$ for 
honeycomb ($t'=0$) is shown in Fig.3. 
While the integral Chern number $c_{\rm F}$ is defined 
only for each gap, we have plotted $c_{\rm F}$ as step functions, 
which makes sense as long as 
the magnetic field is not too large, as in Fig. \ref{fig:3} 
with $\phi=1/31$, since the width of each Landau band 
is then much smaller than the size of gaps.  

The result shows a striking answer to 
a key question: what is the fate of the 
Dirac-like behavior as we go away from $E=0$.  
When the Fermi energy is swept in the honeycomb lattice, 
the Dirac-like Hall conductivity with steps of two
(or four when spin degeneracy is included) 
in units of $-e^2/h$ exist around $E=0$ as has been noted by many 
authors.\cite{Gus05,Nov05,Zha05,SSW06}  Let us call this Dirac-Landau 
behavior.  
 Now, we can see that this Dirac-Landau behavior persists 
{\it all the way up to} some energy,
$E_{vH}=|t|$,
 which we can identify to be 
the van Hove singularities appearing in Fig. \ref{fig:3}.  
We can notice that {\it huge steps} accompanied by a 
sign change in the Hall conductivity occurs at these points.  
This result implies the following: 
The effective theory near the zero energy
is Dirac-like fermions in the continuum limit, for which 
an unconventional quantization of the Hall conductivity has been derived 
in the continuum limit from the Dirac Landau levels.\cite{Gus05,ZhengAndo}  
The present calculation leads to the conclusion that the unusual 
property extends to an unexpectedly wide region of energy 
in the actual lattice fermion system.

Outside the van Hove energies (i.e., in the band-edge regions), 
we recover the conventional quantum Hall effect (QHE) where the step changes 
by one in units of $-e^2/h$ (which we call Fermi-Landau).  
This automatically implies that a huge step accompanied by a 
sign change has to occur at the boundary 
between Dirac-Landau and Fermi-Landau 
behaviors.  In other words 
the bands situated at the van-Hove singularities 
induce a change, which is topological in that the 
relevant quantum numbers are topological. 
The Hall conductivity per spin on 
the honeycomb lattice for the entire energy region is summarized as
( $N =0,1,2,\cdots$ )
\begin{alignat*}{1} 
\sigma_{xy} =& -\frac {e^2}{h} \times
\left\{
\begin{array}{lll}
+(N+1)    & 
E_{\rm F}<-E_{vH},  
&
: 
{\text{ Landau index from the band bottom}}
\\
-(2N+1)    & 
-E_{vH}<E_{\rm F}<0, 
&
:
 { \text{ Dirac-Landau index from the zero energy }}
\\
+(2N+1)    & 
0<E_{\rm F}<+E_{vH}, 
&
:
 { \text{ Dirac-Landau index from the zero energy }}
\\
-(N+1)    & 
E_{vH}<E_{\rm F},
&
: {\text{ Landau index from the band top}}
\end{array}
\right.
\end{alignat*} 
We should again double these numbers if we include the spin degeneracy.
While the unconventional quantization around the zero energy has been
beautifully observed experimentally \cite{Nov05,Zha05}, 
pushing the chemical potential further to approach the van Hove energies 
should detect the topological phase transition.

\section{Bulk-Edge Correspondence in Graphene: Edge states of the Dirac Fermions}

When a system has a nontrivial topological structure, 
edge states should generically exhibit 
characteristic properties. 
For the honeycomb lattice, 
edge states and their flat dispersion 
have been intensively discussed with 
its relevance to spin alignment\cite{Fuji96,Waka98}. 
Another way of saying is that the honeycomb system has a bipartite 
symmetry, 
which guarantees the presence of the zero mode edge states\cite{Ryu02}, i.e., 
there exist macroscopic edge states at the zero energy 
unless the bipartite symmetry is broken.
This gives rise to dispersionless edge states.  
The flat bands are unstable against
 bipartite symmetry breaking (Peierls instability), 
 which is, e.g., realized when an antiferromagnetic spin ordering occurs.  
This may be viewed as a topological origin of the boundary spin moments.\cite{Ryu03cb}

Now, the problem of edge states vs bulk states is particularly 
interesting for the QHE, since the problem of 
how the Dirac-Landau QHE $\propto (2N+1)$ is related to the edge 
states is fundamentally interesting.  
One of the present authors has shown, for 2D 
square (or anisotropic square) systems, that the edge
states whose energy dispersions run across 
the Landau bands have topological QHE numbers, for which 
$
\sigma_{xy}^{\rm edge} =\sigma_{xy}^{\rm bulk}
$
by identifying the connection between the topological integers
for the bulk and for the edge states.\cite{Hatsugai93b,yh-qhe-review}
So let us now focus on the edge states of the honeycomb lattice 
in terms of the topological structure.

\begin{figure}
\begin{center}
\resizebox{0.98\columnwidth}{!}{%
\includegraphics{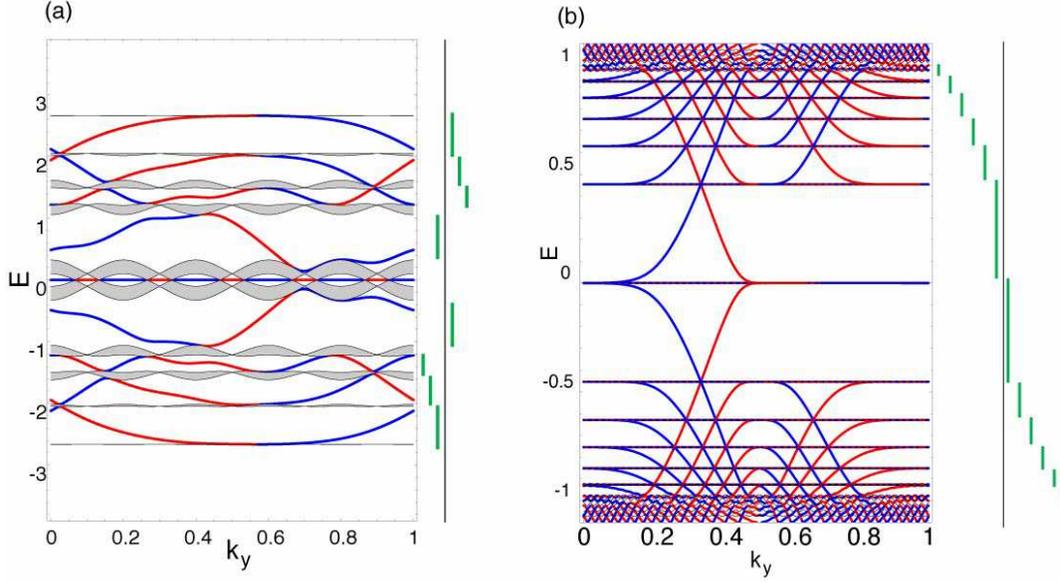} 
}
\caption{
The energy spectra for the cylindrical system in Fig. 
\ref{fig:1}
with the flux $\phi=1/5$ for 
 $t'/t=0$ (honeycomb). 
$k_y$ is a momentum along the cylindrical direction.
Blue lines represent edge states localized at the right (bearded) 
edge, while red ones the left (zigzag) edge, with the bulk energy bands shown as grey regions. 
}
\label{fig:4}       
\end{center}
\end{figure}
We follow the analysis 
in refs\cite{Hatsugai93a,Hatsugai93b,yh-qhe-review} 
to 
reduce the system to a one-dimensional model by 
a partial Fourier transform along the second direction ($j_2)$ 
of the fermion operators,
$
c_{\circ,\bullet}(\mb{j}) =  
L_2^{1/2} \sum_{k_2} e^{ik_2 j_2} c_{\circ,\bullet}(j_1,k_2) 
$.
This yields a $k_2$-dependent series of one-dimensional Hamiltonian 
\begin{alignat*}{1} 
&
H =  \sum_{k_2} 
H_{\rm 1D}(k_2),\nonumber \\
&
H_{\rm 1D}(k_2) =\sum_{j_1}\Big[
t_{\circ\bullet}(j_1,k_2)
c_\circ ^\dagger (j_1,k_2)
c_\bullet  (j_1,k_2) 
  +
t_{\bullet\circ}(j_1,k_2)
c_\bullet ^\dagger  (j_1+1,k_2)
c_\circ   (j_1,k_2)\Big]
+ {\rm h.c.} 
\label{OneDimLocHam}
\end{alignat*} 
where $k_2$-dependent hopping parameters are
\begin{alignat*}{1}
t_{\circ\bullet}(j_1,k_2)=& t\left(1+e^{ik_2-i 2\pi \phi j_1}\right) ,
\nonumber\\
t_{\bullet\circ}(j_1,k_2)=&t\left[1+ (t'/t)~e^{ik_2-i 2\pi \phi
 (j_1+1/2)}\right] .
\end{alignat*} 

We consider 
cylindrical systems with a zigzag edge at one end 
and a bearded (or Klein's) edge at the other. 
We have adopted this geometry, since zigzag-bearded system 
has an $E=0$ edge state in zero magnetic field that 
extend over the entire Brillouin zone, as 
first pointed out in \cite{kusakabe}.

A remarkable feature in the spectrum shown here is that, on top of the edge states 
across adjacent Landau bands, 
there always exist exactly $E=0$ edge states with 
a flat dispersion, which has to exist 
since the number of the edge states are odd, $2q-1$.  
Namely, a system with zigzag edges has 
a zero-energy edge mode in some region of $k_2$, 
a system with bearded edges in the other region, 
and the present system has the mode over the entire region.  
This is a simple extension of the discussion in the 
absence of a magnetic field. \cite{Ryu02,Ryu03cb}

\begin{figure}
\begin{center}
\resizebox{0.78\columnwidth}{!}{%
\includegraphics{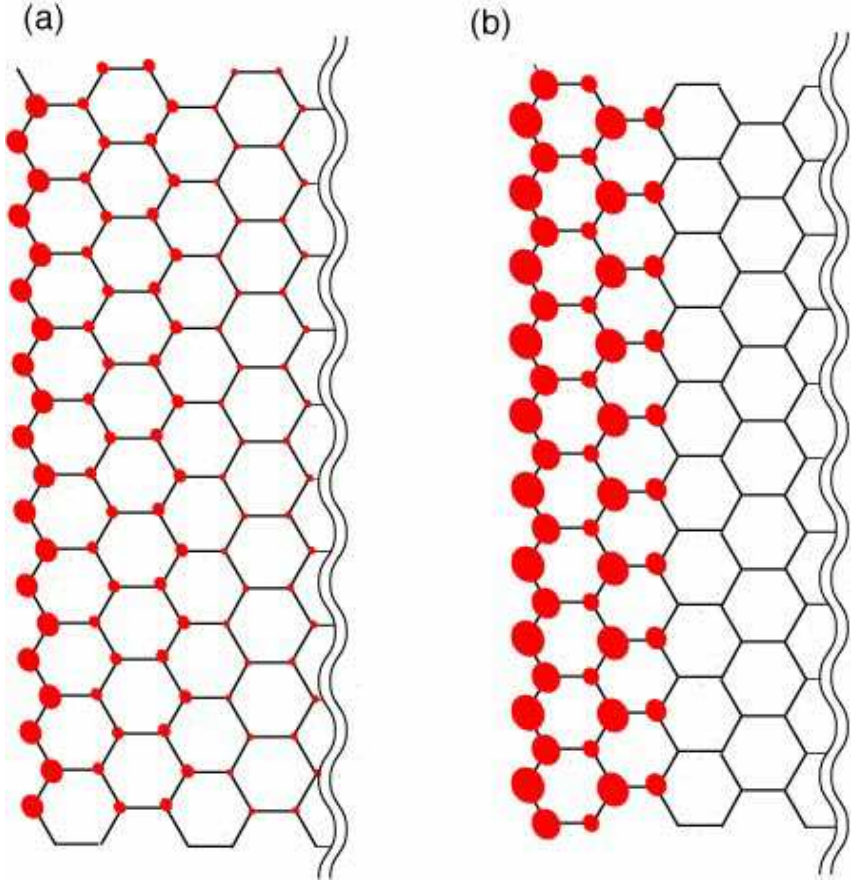} 
}
\caption{
The charge density of the
edge states at the zigzag edge for $\phi=1/3$.
(a)$E=-0.95$ and $E=-1.5$.
}
\label{fig:5}       
\end{center}
\end{figure}

With Laughlin's argument\cite{Laughlin81}, 
supplemented by the 
behavior of the edge state (as a function of the momentum $k_2$), 
one can identify a topological number $I(E_{\rm F})$, 
the number of electrons carried by 
Laughlin's adiabatic procedure, for each edge branch when
the chemical potential $E_{\rm F}$ 
traverses the branch. \cite{Laughlin81,Hatsugai93a,Hatsugai93b,yh-qhe-review}
 As shown in Fig.4, we can readily confirmed a bulk-edge correspondence,
\begin{alignat*}{1} 
{\rm c}_{{\rm F}}(E_{{\rm F}}) &=I(E_{{\rm F}})
\\
\sigma_{xy}^{\rm bulk}
&=
\sigma_{xy}^{\rm edge}
\end{alignat*} 
for Dirac fermions
in graphene.

As for the edge states, we display in Fig. 5 typical 
wave functions in real space.  We can see that 
each of the left and right edge states decays 
into the bulk, where the decay length depends on 
whether its energy is close to the touching 
of the edge mode and the Landau level.

\section{summary}
In summary, we have shown that the Landau levels are divided by the van Hove
singularities into two regimes:
One is effectively described by Dirac particles, and the other by
ordinary finite-mass fermions. 
We have confirmed this both from the 
Chern number argument 
 and 
by the bulk-edge correspondence. 
Topological considerations are useful throughout.  

The properties revealed here for the whole spectrum of the honeycomb lattice 
may be experimentally observable if the chemical potential can be 
varied over a wide range.  
In real graphene samples there may be disorder, in which case 
we are talking about a dirty Hofstadter system.  
However, non-monotonic behaviors should survive the disorder as far 
as the degree of disorder is not too large, as has been 
indicated by a numerical calculation for a dirty Hofstadter system\cite{aoki92}.

\end{document}